\documentclass[pra,aps,superscriptaddress,twocolumn,showpacs,floatfix]{revtex4}
\usepackage{amsmath,amsfonts,amssymb,graphics,graphicx,epsfig,color,times}

\newcommand{\be}{\begin{equation}}
\newcommand{\ee}{\end{equation}}
\newcommand{\ba}{\begin{eqnarray}}
\newcommand{\ea}{\end{eqnarray}}
\newcommand{\ban}{\begin{eqnarray*}}
\newcommand{\ean}{\end{eqnarray*}}

\newcommand{\ket}[1]{\mbox{$ | #1 \rangle $}}

\newcommand{\one}{\leavevmode\hbox{\small1\normalsize\kern-.33em1}}

\begin{document}

\title{Simulation of partial entanglement with non-signaling resources}
\author{Nicolas Brunner}
\author{Nicolas Gisin}
\address{Group of Applied Physics, University of Geneva, Geneva, Switzerland}
\author{Sandu Popescu}
\address{H.H.Wills Physics Laboratory, University of Bristol, Tyndall Avenue, Bristol BS8 1TL, U.K.}
\address{Hewlett-Packard Laboratories, Stoke Gifford, Bristol BS12 6QZ, U.K.}
\author{Valerio Scarani}
\address{Centre for Quantum Technologies and Department of Physics, National University of Singapore, Singapore}
\date{\today}

\begin{abstract}
With the goal of gaining a deeper understanding of quantum
non-locality, we decompose quantum correlations into more
elementary non-local correlations. We show that the correlations of all pure entangled states of two qubits can be simulated without communication, hence using only non-signaling resources. Our simulation model works in two steps. First, we decompose the
quantum correlations into a local and
a non-local part. Second, we present a model for simulating the
nonlocal part using only
non-signaling resources. In our model partially entangled states
require more nonlocal resources than maximally entangled states,
but the less the state is entangled, the less frequently must the
nonlocal resources be used.
\end{abstract}\maketitle

\section{Introduction}

Quantum correlations are very peculiar, especially those violating
some Bell inequality \cite{bell64}. Gaining a
deeper insight into such nonlocal quantum correlation is a grand
challenge. Children gain understanding of how their toys function
by dismantling them into pieces. In the present paper we follow a
similar approach by decomposing the quantum correlations into
simpler, more elementary, nonlocal correlations.

This work is part of the general research program that looks for
nonlocal models compatible and incompatible with quantum
predictions. The goal is to find out what is essential in quantum
correlations. Note that we do not claim that Nature functions as
our model. Nevertheless we believe that finding the minimal
resources sufficient to simulate quantum correlations, and
studying the computational power that they offer
\cite{BHK,NLcrypto,DevIndep,brukner04}, provide enlightening
insights into the quantum world.

%\begin{figure}[b!]
%\begin{center}
%\includegraphics[width=0.55\columnwidth]{prbox.eps}
%\caption{The Popescu-Rohrlich box is the elementary resource
%needed for simulating maximally entangled states of two qubits.
%Note that $\oplus$ means the sum modulo 2.} \label{resources}
%\end{center}
%\end{figure}

In the last years, two different ways of decomposing quantum
correlations have been proposed. The first one, due to Elitzur,
Popescu and Rohrlich (whence EPR-2 !) \cite{EPR2}, consists in
decomposing some quantum correlations into a local and a non-local
part. A second approach consists in the simulation of entanglement
with the help of some non-local resource, e.g. classical
communication or a non-local box. While communication models
\cite{toner,regev} give insight to quantum correlations from the
point of view of communication complexity, we believe that models
using only no-signaling resources \cite{prsinglet} are more
relevant from a physical point of view, since it is most unlikely
that Nature uses any form of communication \cite{beforebefore}. In this paper, we
shall combine for the first time both approaches, and prove that all pure entangled states of two qubits can be simulated using only no-signaling resources, i.e. without communication.

The approach we follow works in two steps. First, in the EPR-2 spirit, we decompose the
quantum correlation $P_Q$ corresponding to von Neumann
measurements performed on pure entangled states of two qubits
$\ket{\psi(\theta)}=\cos{\theta}\ket{00}+\sin{\theta}\ket{11}$
into a statistical mixture of a local correlation $P_L$ and a
non-local correlation $P_{NL}$ \cite{EPR2}:

\ba\label{EPR2} P_{Q} = p_L(\theta) P_L + (1-p_L(\theta))P_{NL}
\,\,.\ea The weight $p_L(\theta)$ is thus a measure of the
locality of the state $\ket{\psi(\theta)}$. In particular, for any maximally entangled state of two-qubits one has $p_L(\theta=\pi/4)=0$ \cite{EPR2}, a result that holds true for maximally entangled states in any dimension \cite{BKP}. Note that in general the probability distribution
$P_{NL}$ does not need to be quantum, but is restricted to
no-signaling correlations by construction.

Then, we provide a
simulation of the nonlocal correlation $P_{NL}$ using only
nonlocal, but non-signaling resources. Accordingly, in order to
simulate $P_Q$, it suffice to simulate $P_L$ with probability
$p_L(\theta)$, which requires only shared randomness (but no nonlocal resources), and
to simulate $P_{NL}$ with the complementary probability
$1-p_L(\theta)$. As expected, the less the quantum
state $\ket{\psi(\theta)}$ is entangled, the smaller the weight
$p_L(\theta)$ of the local correlation \cite{EPR2,SEPR2}. Consequently, the
simulation of a less entangled state requires less frequent use
of nonlocal resources; in particular for separable states $p_L(\theta=0)=1$. However,
not much is known about the nonlocal resources needed to simulate
the nonlocal part of quantum correlations, i.e. to simulate
$p_{NL}$. Reference \cite{prsinglet} presented a simulation of the
quantum correlation for the special case of maximally entangled
qubit pairs ($\theta=\pi/4$) using only one nonlocal box, the so-called PR-box \cite{PR}.
For non-maximally entangled qubit states, very few is known. To our knowledge, the only known result is
that one PR-box is not sufficient for simulating slightly
entangled states \cite{NJP}. This result shows that entanglement and non-locality are different resources, as also suggested by other works \cite{broadbent,eberhard,CabelloDetLoop,AsymDetLoop}.

In this paper we use a decomposition of the form \eqref{EPR2},
recently presented in Ref. \cite{SEPR2}, which is optimal under
some general assumption, and present a simulation of the
corresponding nonlocal correlation $P_{NL}$ for arbitrarily
entangled two qubit states. This simulation requires finitely many
nonlocal boxes, though no claim of optimality can be made. For
pedagogical reasons, the paper is organized as follows. After introducing the general framework in Section 2, and briefly reviewing the case of maximal entanglement in Section 3, we
present in Section 4 a preliminary model for simulating partially entangled
qubit states, without using any decomposition into local and
nonlocal parts. This allows us to introduce the two main ingredients of our model: first the technique of correlated local flips; second the Millionaire box, a generalization of the PR-box. Then in Section 5, we briefly recall the decomposition into local and non-local parts presented in \cite{SEPR2}, and explain how our preliminary simulation model can be extended to
simulate the nonlocal part $P_{NL}$ of the model of Ref. \cite{SEPR2}. Finally we give some conclusions and perspectives.

\section{General framework}

Formally, a correlation is a conditional probability distribution
$P(\alpha\beta | \vec{a}\vec{b})$, where $\alpha$, $\beta$ denote
the outcomes observed by Alice and Bob when they perform
measurements labeled by $\vec{a}$ and $\vec{b}$. Here, measurements are conveniently represented as vectors on the Bloch sphere, since we focus on von Neumann measurements on qubits.
A correlation is non-signalling if and only if Alice and Bob's marginals $M_A$ and $M_B$, are
independent of the partner's input: $M_A$ does not depend on $\vec b$ and $M_B$ does not depend on $\vec a$.
For binary outcomes ($\alpha,\beta \in \{ -1,+1\}$), the
correlations are conveniently written as

\ba P(\alpha,\beta | \vec{a},\vec{b}) = \frac{1}{4} \left( 1 +
\alpha M_A(\vec{a}) + \beta M_B(\vec{b}) + \alpha\beta
C(\vec{a},\vec{b}) \right)  \ea where  \ba\nonumber M_A(\vec{a}) &=& \sum_{\alpha,\beta} \alpha P(\alpha,\beta | \vec{a},\vec{b}) \\ M_B(\vec{b}) &=& \sum_{\alpha,\beta} \beta P(\alpha,\beta | \vec{a},\vec{b}) \,\, , \ea are the local marginals, and
\ba C(\vec{a},\vec{b})=\sum_{\alpha,\beta} \alpha \beta P(\alpha,\beta|\vec{a},\vec{b}) \,\,  \ea is the correlation term. Here we shall focus on pure entangled states of two qubits
$\ket{\psi(\theta)}=\cos{\theta}\ket{00}+\sin{\theta}\ket{11}$, $\theta \in ] 0, \pi/4]$.
Thus the quantum correlation $P_{Q}(\alpha\beta|\vec{a}\vec{b})$
is given by

\ba\nonumber\label{P_Q} M_A(\vec{a}) = ca_z \, \, , \, \, M_B(\vec{b}) = cb_z \, \, \\
 C(\vec{a},\vec{b})= a_z b_z + s(a_xb_x- a_yb_y) \, ,\ea where $c\equiv
 \cos{2\theta}$ and $s\equiv \sin{2\theta}$.

Now, we would like to decompose the correlations $P_Q$ into
simpler ones, such that

\ba P_Q(\alpha,\beta | \vec{a},\vec{b}) = \int d\lambda
P_{\lambda}(\alpha,\beta | \vec{a},\vec{b}) \quad , \ea where
$d\lambda$ is a normalized measure. In a simulation model, two ingredients are required: first, a non-local resources
for creating the elementary correlations $P_{\lambda}$; second, a strategy
(represented by the $\lambda$'s) for judiciously combining them. In this paper, we provide such a
decomposition. The remarkable feature of our model is that the
elementary nonlocal correlations are obtained without
communication, that is using only no-signaling resources.

\section{Maximally entangled state}

Let us briefly the simple case of maximal entanglement, i.e. $\theta=\pi/4$. In this case the marginals vanish, $M_A(\vec a)=M_B(\vec b)=0$,
and the correlation takes the simple scalar product form $C(\vec a,\vec b)=\vec a \cdot \vec b$ \footnote{The sign change in (3) can be adapted locally by one of the parties, for instance by Alice: $a_y \rightarrow -a_y$.}.
In reference \cite{prsinglet} a model simulating this
correlations is presented. This model
uses as resources only shared randomness and one PR-box, which satisfies the relation $a \oplus
b=xy$, where $x$, $y$ are Alice's and Bob's input bits, and $a$,
$b$ their outcome bits. In general nonlocal boxes
provide some elementary nonlocal correlations. They are elementary
in that they allow only for a limited (usually finite) number of
inputs and outputs and they are extremal points in the convex set
of nonsignaling correlations \cite{barrett}. They are nonlocal in
the sense that they violate some Bell inequality. Importantly,
they do not allow signalling, that is the statistics of the local
outcomes (i.e. the marginals) are independent from the other
parties inputs. This model demonstrates that the resource needed
to simulate maximally entangled qubit pairs is surprisingly
simple. Indeed, what could be simpler than $a \oplus b=xy$?

\section{Partially entangled states: preliminary model}

We now turn to partially entangled states of two qubits. In
general the marginals $M_A(\vec a)$ and $M_B(\vec b)$ do not
vanish. However, in the case of two parties and binary outcomes,
it is proven that all extremal nonlocal boxes have vanishing (or deterministic)
marginals \cite{jones,unitPR}. This explains in part why it is difficult
to simulate partially entangled states. In order to circumvent
this difficulty we introduce now the concept of correlated local flips.

\subsection{Correlated local flips.}

Let us consider an arbitrary probability distribution \ba P_0(\alpha, \beta | \vec{a},\vec{b}) = \frac{1}{4} (1 + \alpha\beta
C_0(\vec{a},\vec{b})) \,\, , \ea
with vanishing marginals and correlation term $C_0(\vec{a},\vec{b})$.
Now, Alice and Bob perform local flips on the probability distribution $P_0$; that is, Alice
(Bob) flips her (his) output $-1$ with a probability $f_a$ ($f_b$),
while the output $+1$ is left untouched. After this
processing, also called a Z-channel, the marginals are clearly biased
towards +1. Let us now assume that $f_b \geq f_a$ and that the
flips of Alice and Bob are both determined by a shared random
variable $\Lambda$ uniformly distributed in $[0,1]$. Alice and Bob
flip their -1 outcome if and only if $\Lambda < f_a$ and $\Lambda
< f_b$, respectively. The resulting probability distribution reads

\ba & & P_f (\alpha, \beta | \vec{a}, \vec{b}) = \\\nonumber & &\frac{1}{4} (1 + \alpha f_a + \beta f_b
+ \alpha\beta(f_a + (1 - f_b)C_0(\vec{a}, \vec{b})))\,. \ea It should be pointed out
that the flips $f_a$ and $f_b$ must be correlated; this will be
crucial in the following. Note also that every probability
distribution $P(\alpha, \beta)=\frac{1}{4}(1+\alpha M_A + \beta
M_B + \alpha\beta C)$ with $M_B \geq M_A$ can be generated in this
way.

\subsection{Preliminary model, step 1}

We just described a technique for creating a probability distribution $P_f$
with nontrivial (i.e. non vanishing) marginals, starting from an initial probability
distribution $P_0$ which had trivial marginals. Now the intuition is the
following: since correlation with trivial marginals seem to be
easier to create with standard nonlocal resource (such as PR-boxes),
let us do the identification $P_f = P_Q$ and find out what is the
required initial distribution $P_0$. For partially entangled
states of two-qubits ($P_Q$ given by \eqref{P_Q}), this leads to

\ba f_a=ca_z \,\, , \,\, f_b=cb_z  \,\, ,\,\, C_0 = \vec{a}\cdot
\vec{B} \ea where \ba\label{vecB} \vec{B} \equiv (sb_x, -sb_y, b_z-c)/(1-cb_z)\,\, . \ea Note that $|| \vec{B}|| = 1$. Remarkably, $\vec{B}$ corresponds to
Bob's original measurement setting $\vec{b}$ moved one step back
on the Hardy ladder \cite{Hardy}.

Consequently the problem of simulating correlations originating
from von Neumann measurements on partially entangled states
reduces to the problem of simulating the unbiased probability
distribution \ba P_0 = \frac{1}{4}(1+\alpha\beta
\vec{a}\cdot{\vec{B}})\,\,. \ea Such a "scalar product" correlation can
be reproduced with a single bit of communication \cite{toner} or
with a single PR-box \cite{prsinglet}. However, there is a caveat:
Alice and Bob must know wether $b_z\geq a_z$ (as assumed above) or
if on the contrary $a_z\geq b_z$ ! This is due to the fact that
the local flips must be correlated. Note that in the case $a_z\geq b_z$, the initial probability distribution is given by $P_0 = \frac{1}{4}(1+\alpha\beta \vec{A}\cdot{\vec{b}})$, where $\vec{A}$ is defined similarly to equation \eqref{vecB}.

At first sight it may seem that a resource solving this problem
will lead to signaling, because it would reveal a relationship
between Alice's and Bob's measurements. Remarkably, this is not
the case. Next, we show that a no-signaling (non-local) resource
known as the Millionaire box is exactly the tool we need.

\subsection{The Millionaire box.}

Two millionaires challenge each
other: who is richer ? Since millionaires are in general quite
reluctant to reveal how much money they own, they prefer to use
the Millionaire-box (M-box) \cite{Yao}, a nonlocal two-input two-output non-local box. The two outputs $a$,$b$ are binary, ($a,b\in\{0,1\}$), and are locally random in order to ensure no-signaling. The two inputs
$x,y$ can be chosen in the continuous interval $[0,1$]. The M-box is characterized by the following relation: \ba a \oplus b = [x\leq y] \,\,, \ea where $[X]$ denotes the logical value of $X$:
$[X]=0$ when $X$ is true. Note that the M-box admits an infinite
number of possible inputs. So, both millionaires input the amount of money they own $x$,$y$
into the machine; the parity of the outputs ($a \oplus b$)
indicates the winner. Fortunately, the M-box is also useful to
physicists, as will be shown in the next section. Note that the M-box is a generalization of the PR-box; in
case the inputs $x$, $y$ are binary, the M-box is simply
equivalent to a PR-box (given here by $x(y+1)=a \oplus b \oplus 1 $). It is also worth mentioning that
the M-box reaches the no-signaling bound of all the Bell
inequalities $I_{NN22}$ introduced in \cite{dan}. An interesting question is whether all (bipartite) non-local boxes with two-outcomes \cite{jones,unitPR} can be simulated with one M-box. Indeed, a detailed study of the non-local
properties of the M-box would be relevant, but is beyond the scope of this paper.

\subsection{Preliminary model, step 2}

As shown above, the technique of local flips allows one to recover the
correlation of partially entangled states, under the condition
that $b_z\geq a_z$ (or $a_z\geq b_z$). But how do Alice and Bob
know whether $b_z \geq a_z$ or $a_z \geq b_z$ ? The M-box can overcome this problem.

\begin{figure}[t!]
\begin{center}
\includegraphics[width=0.9\columnwidth]{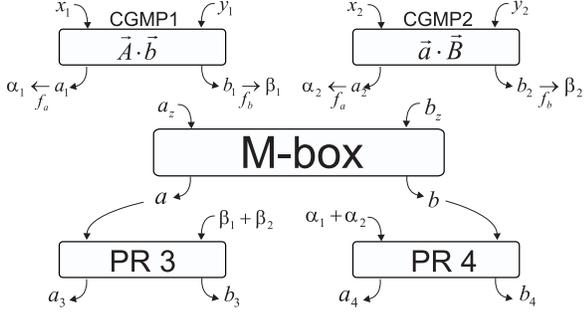}
\caption{Preliminary model. Simulating partial entanglement without communication. The model requires four PR-boxes and a Millionaire-box (M-box). The first two PR-boxes create "scalar product" correlations (CGMP-boxes). Then the M-box "selects" the correct CGMP-box, without revealing any relation between Alice's and Bob's measurement settings (i.e. without signaling). Finally, two additional PR-boxes are required for computing the correct outputs.} \label{Model1}
\end{center}
\end{figure}

Alice and Bob share two PR-boxes for creating "scalar product"
correlations (see Fig. \ref{Model1}); from now on we call these
CGMP-boxes \cite{prsinglet}. The first one is used to create the
correlation given by the scalar product $\vec{a}\cdot\vec{B}$,
i.e. corresponding to the case $b_z\geq a_z$ and the second one
for the scalar product $\vec{A}\cdot\vec{b}$, i.e. for the case
$a_z\geq b_z$. Local flips are then performed. At this point,
Alice and Bob have each got two possible outputs $\alpha_1,\alpha_2$ and
$\beta_1,\beta_2$, but don't know which one to use, since they don't know whether $a_z\le b_z$ or $b_z\le a_z$.

Next, they input the $z$-component of their measurement setting
(respectively $a_z$ and $b_z$ \footnote{Since $P_Q (-\alpha,\beta
| -\vec{a},\vec{b})=P_Q (\alpha,\beta | \vec{a},\vec{b})$, it is
sufficient to consider the case where $a_z,b_z \geq 0$.}) into the
M-box, and get outputs $a$ and $b$. It is clear that, for the
simulation to succeed, the final output of Alice and Bob, $\alpha$ and
$\beta$, should be equal to $\alpha_1$,$\beta_1$ if $a_z \geq b_z$, and equal
to $\alpha_2$,$\beta_2$ if $b_z \geq a_z$. Mathematically this translates
into the following expression

\ba\label{OUT} \alpha \oplus \beta = (a \oplus b)(\alpha_1 \oplus  \beta_1) \oplus (a \oplus b \oplus1)(\alpha_2 \oplus \beta_2) \, .\ea
Developing the previous equation, one gets

\ba\nonumber \alpha \oplus \beta &=& a(\alpha_1 \oplus \alpha_2) \oplus  \alpha_2  \oplus
b(\beta_1  \oplus \beta_2) \oplus \beta_2 \\\label{outputs} & & \oplus a(\beta_1 \oplus \beta_2) \oplus
b(\alpha_1 \oplus \alpha_2) \, ,\ea which contains some local terms, as well
as some non-local terms. Remarkably, the non-local terms (second line of equation \eqref{outputs}) are simply obtained by using two supplementary
PR-boxes, $a_3 \oplus b_3 = a(\beta_1 \oplus \beta_2)$, and $a_4 \oplus
b_4 = b(\alpha_1 \oplus \alpha_2)$ (see Fig. \ref{Model1}).

So finally, using four PR-boxes (two CGMP-boxes and two additional
PR-boxes) and one M-box, one can simulate the correlation of any
partially entangled state of two qubits. Whether the M-box can be
replaced by a finite number of PR-boxes (or more generally with a
nonlocal box having a finite number of possible inputs) is an
interesting open question.

\section{Partially entangled states: main model, integrating EPR-2}

We are now ready to present our model, combining the preliminary model (presented in the previous section) and the decomposition of Ref \cite{SEPR2}, into local and non-local parts (i.e. of the form \eqref{EPR2}). The decomposition is the following:

\ba\nonumber p_L(\theta) &=& 1-s \\\label{SEPR2}  P_L &=& \frac{1}{4} \left( 1+ \alpha f(a_z)  \right)\left( 1+ \beta f(b_z)  \right)  \\\nonumber
P_{NL} &=& \frac{1}{4} \left( 1 +  \alpha F(a_z) + \beta F(b_z) + \alpha \beta G(\vec{a}\vec{b}) \right)  \ea where $f(x) = \textrm{sgn}(x) \textrm{min}(1,\frac{c}{1-s}|x|)$ , $F(x)=\frac{1}{s} (cx-(1-s)f(x))$ , and $G(\vec{a}\vec{b}) = a_x b_x -a_yb_y + \frac{1}{s} [a_z b_z - (1-s)f(a_z)f(b_z)]$ . We refer the reader to \cite{SEPR2} for further details.

Let us point out two important features of decomposition \eqref{SEPR2}  First, the weight of the local part $p_L(\theta)=1-s$ is a
monotonic decreasing function of $\theta$, i.e. of the degree of entanglement of the state $\ket{\psi(\theta)}$. Note also that $p_L(\theta)=1-s$ is optimal under the assumption that $P_L$ depends only on $a_z$ and $b_z$. Second, the non-local part
$P_{NL}$ depends on the measurement settings. More
precisely, when the measurement setting of Alice is such that $a_z\leq
(1-s)/c$ (i.e. inside a slice of the Bloch sphere around the
equator), her local marginal vanishes; and similarly for Bob. On
the contrary, when the measurement setting lies outside the slice,
the marginal is biased. When both the settings of Alice and Bob
are found inside the slice, the correlation reduces to a simple
scalar product with trivial marginals.

\begin{figure}[t]
\begin{center}
\includegraphics[width=0.9\columnwidth]{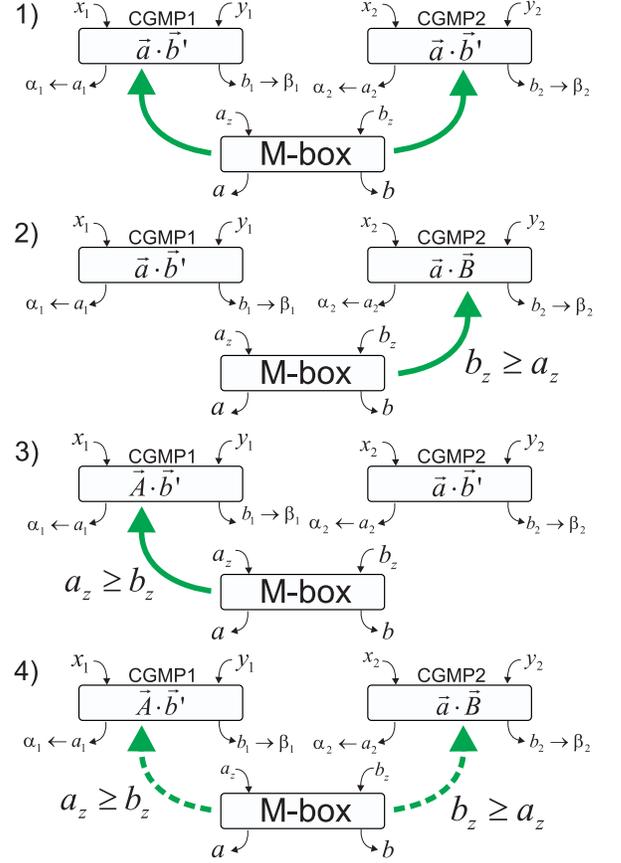}
\caption{Sketch of the main model. 1) Alice and Bob inside the slice (see text);
both CGMP-boxes can be used. 2) Alice inside, Bob outside; then
indeed $a_z\leq b_z$. 3) Alice outside, Bob inside. 4) Alice and
Bob outside. Depending wether $a_z\leq b_z$ or $a_z \geq b_z$, the
M-box selects the correct CGMP-box. Note that we have omitted the
two additional PR-boxes ($PR_3$, $PR_4$ of Fig. 2).} \label{Simul}
\end{center}
\end{figure}

The simulation of $P_{NL}$ is very similar to that presented above, thus
we only describe Alice's and Bob's strategies. As previously, the
required non-local resources are two CGMP-boxes, an M-box and two
additional PR-boxes. After establishing non-local correlations
with both CGMP-boxes, Alice and Bob perform local flips. Finally
they use two additional PR-boxes to compute the correct output (see Fig. \ref{Simul}).

Alice proceeds as follows. When her setting is inside the slice
($a_z\leq (1-s)/c$), she inputs according to $\vec{a}$ into both
CGMP-boxes, and does not perform any local flip ($f_a=0$). When
her setting is outside the slice, she inputs the first CGMP-box
according to $\vec{A}=(sa_x,sa_y,c-a_z)/(1-a_z c)$ and the second
CGMP-box according to $\vec{a}$. Then she biases her output
towards outcome +1 with probability $f_a = F(a_z)$.

%Note that $\vec{A}$ is again (up to one sign)
%the original setting $\vec{a}$ moved back one step on the Hardy
%ladder.

Bob proceeds almost similarly. When his measurement setting is inside the
slice ($b_z\leq (1-s)/c$), he inputs both CGMP-boxes according to
$\vec{b}' = (b_x,-b_y,-b_z)$. When his setting is outside the
slice, he inputs the first CGMP-box according to $\vec{b}' $ and
the second according to $\vec{B}=(sb_x,-sb_y,b_z-c)/(1-b_z c)$.
Then he biases his output with probability $f_b=F(b_z)$.

\section{Conclusion and Outlook}

By dismantling the quantum correlations of partially entangled
states of two-qubits into more elementary nonlocal but
no-signaling correlations, we gained insight into the quantum
world. We showed that the correlations of all pure entangled states of two qubits (under von Neumann measurements) can be simulated using only non-signaling resources, hence without communication. Our decomposition is likely not to be optimal in the sense
that there might exist more economical models. Still, there are
already two lessons we learn from the present decomposition.
First, the less the quantum state is entangled, the less
frequently one needs to use nonlocal resources to simulate it; as
intuition suggests. Next, whenever one needs nonlocal resources,
then these are definitively larger for (at least some) partially
entangled states than for the maximally entangled state; indeed
this is proven for slightly entangled states \cite{NJP}, but is
still an open question for close to maximally entangled states. Hence, in counting the
resources required to simulate two-qubit states, one
should distinguish between the required amount of nonlocal
resources and the frequency at which one has to use them.

It is interesting to establish the following connection with
Leggett's approach to quantum correlation \cite{Leggett03}, which
recently attracted quite some attention
\cite{LeggettVienna1,LeggettVienna2,LeggettGeneva1,LeggettGeneva2,LeggettSuarez,LeggettRenato}.
In models \textit{\`a la Leggett} one assumes that the elementary
correlations, contrary to PR-boxes, have nontrivial marginals;
Leggett's original idea is that each qubit, when analysed
individually, appears to be always in a pure state, see
\cite{Leggett03,LeggettGeneva2}. However, one can prove that any
such model, with elementary correlation having nontrivial
marginals, fails to reproduce the quantum correlation of maximally
entangled states of two-qubits \cite{LeggettGeneva2,LeggettRenato}. This is a kind of converse to the present paper
in which we show that it is especially hard to simulated at the same time nonlocal correlations and non-vanishing marginals.

Among the open questions, we like to underline the following one.
How could one prove that a decomposition is minimal? As said,
this question has two sides. Minimality of the
resources, and minimality of the frequency at which one has to use
them. Our experience suggests that the first aspect is an
especially difficult problem. The second aspect looks more
promising: it seems natural to conjecture that an EPR2-type
decomposition with $p_{NL}(\theta)=1-\cos{2\theta}$ should exist \cite{SEPR2}.

\textit{Acknowledgments} Fruitful discussions with Harry Buhrman's group are
acknowledged. N.B. and N.G acknowledge financial support from the EU project QAP
(IST-FET FP6-015848) and Swiss NCCR Quantum Photonics; V.S. acknowledges financial support from the National Research Foundation and Ministry of
Education, Singapore. N.B. acknowledges the hospitality of the National University of
Singapore.

\bibliographystyle{prsty}
%\bibliography{H:/BIB/thesis}
\bibliography{C:/BIB/thesis}

\end{document}